\documentclass[usenatbib]{mn2e}
\usepackage{graphicx,natbib,times}
\usepackage{bm,url}
\graphicspath{{./fig/}{./png/}}
\usepackage{color}

\newcommand{\EQ}{\begin{equation}}
\newcommand{\EN}{\end{equation}}
\newcommand{\EQA}{\begin{eqnarray}}
\newcommand{\ENA}{\end{eqnarray}}

\newcommand{\Eq}[1]{Equation~(\ref{#1})}
\newcommand{\Eqs}[2]{Equations~(\ref{#1}) and~(\ref{#2})}

\newcommand{\Fig}[1]{Figure~\ref{#1}}

{}
{}
{}
\newcommand{\meanemf}{\overline{\cal E} {}}

{}
{}
\newcommand{\meanEMF}{\overline{\mbox{\boldmath ${\cal E}$}}{}}{}
{}
{}
{}
{}
{}
{}
\newcommand{\meanBB}{\overline{\mbox{\boldmath $B$}}{}}{}
{}
{}
{}
{}
{}
{}
{}
{}
\newcommand{\meanJJ}{\overline{\mbox{\boldmath $J$}}{}}{}
{}
\newcommand{\meanUU}{\overline{\mbox{\boldmath $U$}}{}}{}

{}
{}
{}

\newcommand{\meanB}{\overline{B}}

\newcommand{\meanJ}{\overline{J}}

{}

{}
{}

\newcommand{\tildeB}{\tilde{B}}
\newcommand{\tildeJ}{\tilde{J}}
\newcommand{\tildeemf}{\tilde{\cal E}}

\newcommand{\bb}{\bm{b}}
\newcommand{\BB}{\bm{B}}

\newcommand{\uu}{\mbox{\boldmath $u$} {}}
\newcommand{\UU}{\mbox{\boldmath $U$} {}}

\newcommand{\JJ}{\mbox{\boldmath $J$} {}}

\newcommand{\WW}{\mbox{\boldmath $W$} {}}

\newcommand{\nab}{\mbox{\boldmath $\nabla$} {}}

\newcommand{\ii}{{\rm i}}

\newcommand{\dd}{{\rm d} {}}

\def\la{\mathrel{\mathchoice {\vcenter{\offinterlineskip\halign{\hfil
$\displaystyle##$\hfil\cr<\cr\sim\cr}}}
{\vcenter{\offinterlineskip\halign{\hfil$\textstyle##$\hfil\cr<\cr\sim\cr}}}
{\vcenter{\offinterlineskip\halign{\hfil$\scriptstyle##$\hfil\cr<\cr\sim\cr}}}
{\vcenter{\offinterlineskip\halign{\hfil$\scriptscriptstyle##$\hfil\cr<\cr\sim\cr}}}}}
\def\ga{\mathrel{\mathchoice {\vcenter{\offinterlineskip\halign{\hfil
$\displaystyle##$\hfil\cr>\cr\sim\cr}}}
{\vcenter{\offinterlineskip\halign{\hfil$\textstyle##$\hfil\cr>\cr\sim\cr}}}
{\vcenter{\offinterlineskip\halign{\hfil$\scriptstyle##$\hfil\cr>\cr\sim\cr}}}
{\vcenter{\offinterlineskip\halign{\hfil$\scriptscriptstyle##$\hfil\cr>\cr\sim\cr}}}}}
\def\Rm{R_{\rm m}}

\def\Rey{\mbox{\rm Re}}
\def\Pe{\mbox{\rm Pe}}

\def\kf{k_{\rm f}}

\def\urms{u_{\rm rms}}
\def\wrms{w_{\rm rms}}

\def\kappat{\kappa_{\rm t}}

\def\etat{\eta_{\rm t}}

\def\half{{\textstyle{1\over2}}}

\newcommand{\yapj}[3]{ #1, {ApJ,} {#2}, #3}

\newcommand{\yan}[3]{ #1, {Astron.\ Nachr.,} {#2}, #3}

\newcommand{\yana}[3]{ #1, {A\&A,} {#2}, #3}

\newcommand{\ygafd}[3]{ #1, {Geophys.\ Astrophys.\ Fluid Dyn.,} {#2}, #3}

\newcommand{\yptrs}[3]{ #1, {Phil.\ Trans.\ R.\ Soc.,} {#2}, #3}

\newcommand{\ymn}[3]{ #1, {MNRAS,} {#2}, #3}

\newcommand{\ypre}[3]{ #1, {Phys.\ Rev.\ E,} {#2}, #3}

\newcommand{\yjour}[4]{ #1, {#2}, {#3}, #4}

\newcommand{\ybook}[3]{ #1, {#2} (#3)}
\newcommand{\yproc}[5]{ #1, in {#3}, ed.\ #4 (#5), #2}

\title[Negative eddy diffusivity dynamo]{A mean field dynamo from negative eddy diffusivity}
\author[E.\ Devlen et al.]{Ebru Devlen$^{1,2}$\thanks{E-mail:devlen@nordita.org},
Axel Brandenburg$^{1,3}$ and Dhrubaditya Mitra$^1$\\
$^1$Nordita, KTH Royal Institute of Technology and Stockholm University,
Roslagstullsbacken 23, 10691 Stockholm, Sweden\\
$^2$Department of Astronomy \& Space Sciences, Faculty of Science,
University of Ege, Bornova 35100, Izmir, Turkey\\
$^3$Department of Astronomy, Stockholm University, 10691 Stockholm, Sweden}

\date{\today,~ $ $Revision: 1.63 $ $}
\begin{document}
\maketitle

\label{firstpage}

\begin{abstract}
Using direct numerical simulations, we verify that ``flow IV''
of Roberts (1972) exhibits dynamo action dominated by horizontally
averaged large-scale magnetic field.
With the test-field method we compute the turbulent magnetic diffusivity
and find that it is negative and overcomes the molecular diffusivity,
thus explaining quantitatively the
large-scale dynamo for magnetic Reynolds numbers above $\approx8$.
As expected for a dynamo of this type, but contrary to
$\alpha$-effect dynamos, the two horizontal field
components grow independently of each other and have arbitrary
amplitude ratios and phase differences.
Small length scales of the mean magnetic field are shown to be stabilized
by the turbulent magnetic diffusivity becoming positive at larger wavenumbers.
Oscillatory decaying or growing solutions have also been found in certain
wavenumber intervals and sufficiently large values of the magnetic
Reynolds number.
For magnetic Reynolds numbers below $\approx0.5$ the turbulent magnetic
diffusivity is confirmed to be positive, as expected for all incompressible
flows.
Earlier claims of a dynamo driven by a modified Taylor-Green flow
through negative eddy diffusivity could not be confirmed.
\end{abstract}


\section{Introduction}

The equations of magnetohydrodynamics (MHD) permit growth of magnetic energy at the expense of kinetic
energy.
This phenomenon is called the dynamo effect \cite[see e.g.][for a recent review]{BS05}.
If the dynamo effect gives rise to a magnetic field whose
characteristic length scale is greater than that of the fluid, we call
it a {\it large-scale} dynamo.
Most astrophysical dynamos, including the solar dynamo and the galactic dynamo, are of this
type.

To theoretically describe the large-scale dynamo one must average the equations of MHD over the
small scales to obtain an effective equation for the large-scale magnetic field.
This effective equation can be written down by
using either mean-field theory \citep{SKR66} or multiple-scale
expansions \citep[see e.g.][for a recent review]{Zhe12}.
These equations contain turbulent transport coefficients: the $\alpha$
effect and turbulent diffusivity.
In general, both are tensors whose complexity depends on the symmetries of the problem.
Within the formalism of mean-field theory, it is generally a non-trivial task to calculate the turbulent transport
coefficients even if we ignore the back-reaction of the magnetic field on the flow, i.e., for kinematic dynamos.
For several kinematic problems, the turbulent transport coefficients have
been calculated using the test-field method (TFM) of \cite{Sch05,Sch07};
see also \cite{B05_QPO,BRRK08,BRS08}.
Typically, it is found that the $\alpha$ effect gives rise to
the growth of a large-scale magnetic field while
the turbulent diffusivity contributes to decay by effectively enhancing
the molecular magnetic diffusivity.
However, multiscale methods have shown that for certain flows the $\alpha$
effect can be zero, but the eddy diffusivity, i.e., the sum of turbulent
and molecular diffusivity, may turn out to be
negative \cite[see e.g.,][]{LNVW99,ZPF01}.
In that case, such flows may act as large-scale dynamos.
However, we are not aware of direct numerical simulations (DNS)
that demonstrate that those flows really do produce mean magnetic fields
and that this is caused by negative eddy diffusivity.

In a remarkable paper, \cite{Rob72} shows that the multiple-scale
versions of two-dimensional spatially periodic motions can give growing
magnetic fields for magnetic diffusivities below a critical value.
He studies four different periodic flow patterns.
We are here especially interested in Flow IV
(in the following referred to as Roberts-IV flow) because,
although this flow yields exponentially growing solutions in time
(see his Figure~10), he finds all components of the $\alpha$ tensor
to be zero.
\cite{Rob72} also notes that his results are relevant to turbulent
dynamos with positive turbulent diffusivity, but the possibility of
negative turbulent diffusivity is not discussed explicitly.
In this paper we first verify, using DNS,
that for a particular flow \citep{Rob72},
namely the Roberts-IV flow, it is possible to drive a
kinematic large-scale dynamo, although the $\alpha$ effect and the
planar-averaged kinetic helicity are indeed zero.
Next, by using the TFM, we show that such a dynamo can
be accurately described by zero $\alpha$ effect but negative turbulent
diffusivity which dominates over the molecular one.
(In the context of laminar flows, the expression {\it turbulent}
diffusivity is not optimal, and refers simply to a diffusion-like
coefficient in the averaged equations.)

Finally, we turn to the Taylor--Green and the modified Taylor--Green flows,
for which negative eddy diffusivity dynamos have been claimed previously
\citep{LNVW99}.
Again, these flows have no net helicity.
Although dynamo action was found in several cases, no large-scale
magnetic field was found in DNS of these flows.
Furthermore, the $\alpha$ effect turns out to be zero, but the
eddy diffusivity remains positive.
This flow does therefore not appear to be an example of a
negative eddy diffusivity dynamo.

\section{The Roberts-IV flow}

In connection with understanding the geodynamo, \cite{Til04} studied in
some detail the Roberts IV flow.
We follow here Tilgner's definition of the flow:
\begin{equation}
\UU=u_0\pmatrix{
\sqrt{2/f}\,\sin k_0x\cos k_0y \cr
-\sqrt{2/f}\,\cos k_0x\sin k_0y \cr
\sqrt{f}\,\sin{k_0x}
},
\end{equation}
where $u_0$ characterizes the amplitude of the flow.
It is solenoidal and its vorticity, $\WW=\nab\times\UU$, is given by
\begin{equation}
\WW=u_0\pmatrix{
0\cr
-\sqrt{f}\,k_0\cos{k_0x}\cr
2\sqrt{2/f}\,k_0\sin k_0x\sin k_0y
}.
\end{equation}
Here, the parameter $f$ determines the relative importance of
vertical to horizontal motions.
The kinetic helicity density, $\WW\cdot\UU$, is given by
\begin{equation}
\WW\cdot\UU=\sqrt{2}\,u_0^2 k_0\,(1+\sin\!^2{k_0x})\sin k_0y
\end{equation}
and is independent of $f$.
\cite{Til04} showed that in spite of the horizontally averaged kinetic helicity density,
$\overline{\WW\cdot\UU}$, being zero; the Roberts IV flow gives rise to dynamo action.
In other words, it leads to growing solutions of the induction equation,
\begin{equation}
{\partial\BB\over\partial t}=\nab\times\left(\UU\times\BB-\eta\JJ\right),
\label{dBdt}
\end{equation}
where $\eta$ is the microphysical (molecular) magnetic diffusivity,
$\BB$ is the magnetic field, $\JJ=\nab\times\BB$ is the
current density, and we have chosen our units such
that the vacuum permeability is unity.

Note, however, that \cite{Til04} described the dynamo to be a small-scale
one, i.e., the characteristic length scales of the magnetic field is of
the same order as $1/k_0$.
In the following, we obtain solutions to \Eq{dBdt} via DNS using
the {\sc Pencil Code}\footnote{\url{http://pencil-code.googlecode.com/}}.
We do not evolve the flow, hence we study kinematic dynamo solutions.

\begin{figure}\begin{center}
\includegraphics[width=\columnwidth]{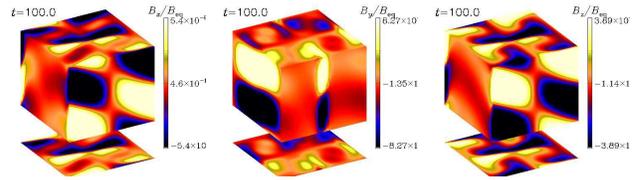}
\end{center}\caption[]{
Three components of the magnetic field on the periphery of the computational
domain for $\Rm=20$, $f=1$, and domain size $L_x=L_y=L_z=2\pi/k_0$.
}\label{B}\end{figure}

\begin{figure}\begin{center}
\includegraphics[width=\columnwidth]{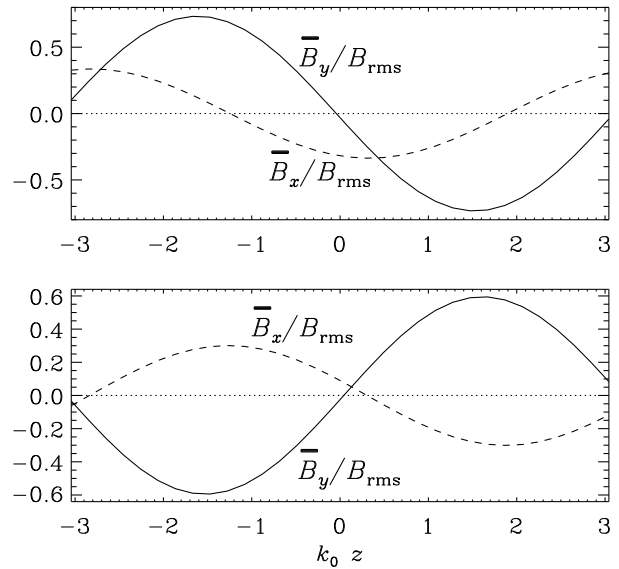}
\end{center}\caption[]{
Examples of runs with two different initial conditions
(upper and lower panels) showing the $x$ and
$y$ components of the mean field (normalized by the rms value of the
total field $\BB$) versus $z$, obtained from the DNS for $f=1$,
$L_x=L_y=L_z=2\pi/k_0$, and $\eta=0.05\,u_0/k_0$, corresponding to $\Rm=20$.
}\label{ppxyaver}\end{figure}

As an example of the resulting magnetic field, we show in \Fig{B}
the three components of the magnetic field at the periphery of the
computational domain.
It is remarkable that the resulting magnetic field has a large-scale
component that survives $xy$ averaging (denoted by overbars), i.e., $\meanBB=\meanBB(z,t)$
is non-vanishing; see \Fig{ppxyaver}, where we show examples of the
resulting mean field obtained by averaging the solution of the DNS.
In other words, we have here an example of a mean-field dynamo with
$\meanBB$ being a solution of the horizontally averaged induction equation,
\begin{equation}
{\partial\meanBB\over\partial t}=\nab\times\left(\meanUU\times\meanBB
+\meanEMF-\eta\meanJJ\right),
\label{dmeanBdt}
\end{equation}
where $\meanEMF=\overline{\uu\times\bb}$ is the mean electromotive force
resulting from correlations of residual velocity and magnetic fields,
$\uu=\UU-\meanUU$ and $\bb=\BB-\meanBB$, respectively.
(Note that here $\meanUU=0$.)
Empirically, we find that the horizontally averaged solutions of
\Eq{dBdt} are of the form
\begin{equation}
\meanBB(z,t)=\pmatrix{
B_{0x}\cos(kz+\phi_x)\cr
B_{0y}\cos(kz+\phi_y)\cr
0}\,e^{\lambda t},
\end{equation}
where $B_{0x}$, $B_{0y}$, $\phi_x$, and $\phi_y$ are arbitrary constants,
i.e., the $x$ and $y$ components of the magnetic field evolve independently
of each other and they have arbitrary phase shifts, depending just on the
properties of the initial conditions; see \Fig{ppxyaver} for an example.
The same result can be inferred from Equation~(7.3) of \cite{Rob72}.
Solutions of \Eq{dmeanBdt} can be obtained by mean-field simulations (MFS)
which requires a closed expression for $\meanEMF$ in terms of $\meanBB$.
This will be discussed in the following.

As pointed out by \cite{Til04}, the Roberts-IV flow has no $\alpha$ effect.
As this is a laminar flow, driving a dynamo via fluctuations of the $\alpha$
effect \citep{MB12} is also not possible.
This suggests that the observed mean field might be produced by a
negative eddy diffusivity \citep{LNVW99,ZPF01}.
To investigate such a possibility, we now apply the TFM to calculate
the turbulent transport coefficients of the Roberts IV flow.

As is long recognized \citep{Rad76}, the connection between $\meanEMF$
and $\meanBB$ is a nonlocal one that is described by a convolution of the form
\begin{equation}
\meanemf_i=\hat\alpha_{ij}\circ\meanB_j-\hat\eta_{ij}\circ\meanJ_j,
\label{meanemfi}
\end{equation}
where ``$\circ$'' denotes a convolution in space and time, i.e.,
\begin{equation}
\hat\eta_{ij}\circ\meanJ_j\!=\! \int\!\!\int
\hat\eta_{ij}(z-z',t-t')\,\meanJ_j(z',t')\, \dd z' \, \dd t',
\end{equation}
and likewise for $\hat\alpha_{ij}\circ\meanB_j$, but this term is
vanishing for the Roberts-IV flow \citep{Til04}.
The hats on $\hat\alpha_{ij}$ and $\hat\eta_{ij}$ indicate that the
corresponding quantities are integral kernels.

We emphasize that in \Eq{meanemfi} we have made use of the fact that
the only non-vanishing derivatives of a horizontally averaged mean field
are $\partial\meanB_x/\partial z$ and $\partial\meanB_y/\partial z$, which
be expressed in terms of components of $\meanJJ$, so the corresponding
turbulent diffusivity tensor is just of rank 2, not, as
in the general case of rank 3 \citep{KR80}.

In the TFM, the kernel formulation above is most naturally considered
in Fourier space with
\begin{equation}
\tildeemf(k,\omega)=
\tilde\alpha_{ij}(k,\omega)\tildeB_j(k,\omega)
-\tilde\eta_{ij}(k,\omega)\tildeJ_j(k,\omega),
\label{EMFkomega}
\end{equation}
where tildes denote appropriately normalized Fourier transforms
of the corresponding mean-field quantities \citep{BRS08,HB09}.
The usual $\alpha$ effect and turbulent diffusivity emerge in the
limits $k\to0$ and $\omega\to0$ for the respective quantities.
Hereafter, we drop the tildes even when the $k$ and $\omega$ arguments
are indicated to be non-vanishing.

In the following we consider a three-dimensional domain of size
$L_x\times L_y\times L_z$.
For most cases we choose cubic domains, i.e., $L_x=L_y=L_z=2\pi/k_0$.
We are primarily interested in the case of harmonic solutions of \Eq{dmeanBdt}
with given vertical wavenumber $k$ of a magnetic field that
is growing or decaying exponentially proportional to $e^{\lambda t}$.
So we are interested in the case $\omega=\ii\lambda$.
As an approximation, we begin by considering the case $\omega=0$,
i.e., we ignore the so-called memory effect; see \cite{HB09} for
illustrating the departure in the case of the standard Roberts flow
with helicity (also known as the Roberts-I flow).

\section{Results}

In the following we use the TFM, as described in \cite{BRS08}
and \cite{HB09}.

\subsection{Sign change of eddy diffusivity}

As we have already mentioned, all components of
$\alpha_{ij}(k,\omega)$ vanish for the Roberts-IV flow.
Moreover, $\eta_{ij}(k,\omega)$
is isotropic, i.e., we can write $\eta_{ij}=\etat\delta_{ij}$.
In practice, we compute $\etat=(\eta_{11}+\eta_{22})/2$ and
find that $\epsilon_\eta=(\eta_{11}-\eta_{22})/2$ vanishes
to numerical accuracy.

A priori the fact that $\eta_{ij}(k,\omega)$ is isotropic is
surprising because the $z$ component of the flow is not isotropic.
This is also confirmed by analytical calculation using the
second order correlation approximation (SOCA), as shown by
\cite{Rad13}.
Indeed, this isotropy is broken once we allow for averages
that depend on $y$ and $z$, but with that definition of averages,
$\alpha$ is no longer zero.
This leads to other interesting interpretations regarding
negative eddy diffusivity dynamos that will be investigated
in a future publication.

\begin{figure}\begin{center}
\includegraphics[width=\columnwidth]{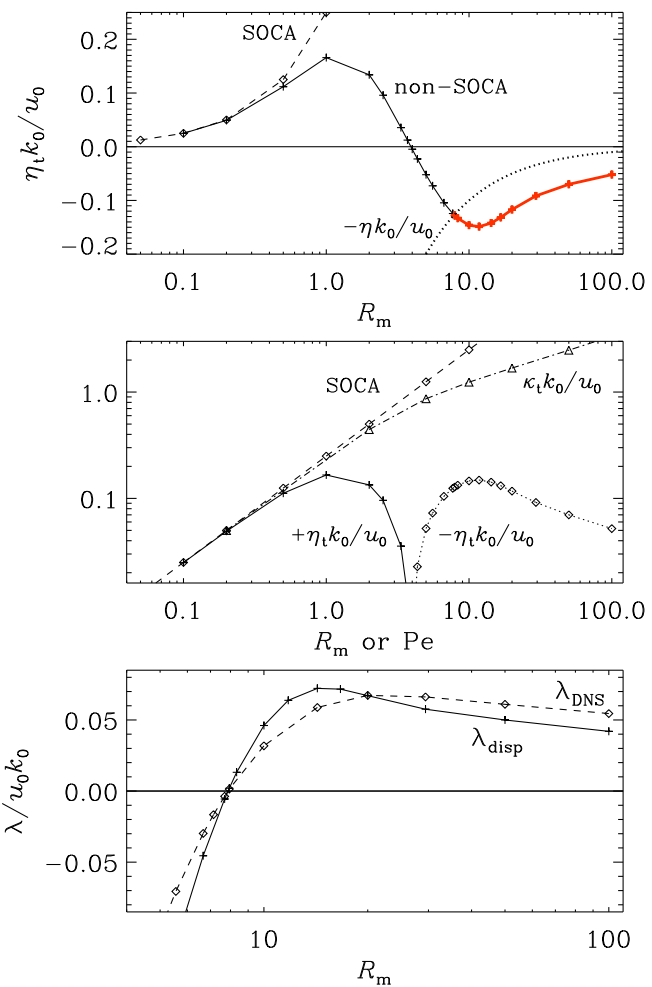}
\end{center}\caption[]{
Turbulent magnetic diffusivity, $\etat$ (top and middle panels), and
growth rates $\lambda_{\rm disp}$ and $\lambda_{\rm DNS}$
versus $\Rm$ for $f=1$ and $k=k_0$.
In the first two panels, the dashed lines give the SOCA result,
$\etat k_0/u_0=\Rm/4$.
In the first panel, the intersection between $\etat$ and $-\eta$
(dotted line) marks the onset of dynamo action at $\Rm\approx8$.
(The section for $\Rm>8$ is marked in red/thick.)
The double-logarithmic representation in the middle panel allows one
to see that the linear SOCA dependence is obeyed for $\Rm\la0.5$.
Note also that the turbulent passive scalar diffusivity, $\kappat$,
remains positive (triangles and dash-dotted line).
}\label{petat}\end{figure}

The resulting values of $\etat(k_0,0)$ are shown in \Fig{petat}
as a function of magnetic Reynolds number,
\begin{equation}
\Rm=u_0/\eta k_0.
\end{equation}
For comparison with earlier work involving turbulent flows,
we note that this definition of $\Rm$
is close to a definition in terms of the rms velocity of the flow (for $f=1$ we
have $\urms\approx1.225\,u_0$) and the wavenumber of the energy-carrying
eddies $\kf$, i.e., $\urms/\eta\kf$.
If we approximate $\kf\approx\wrms/\urms$, where $\wrms$ is the
rms value of the fluctuating part of the vorticity, then
we have $\kf\approx1.29\,k_0$.
Therefore, we have $\urms/\eta\kf\approx0.95\,\Rm$, which is close to $\Rm$.

As is common to many turbulent transport coefficients \citep{SBS08,BRS08},
$\etat$ grows linearly with $\Rm$ for $\Rm\la0.5$; see the
middle panel of \Fig{petat}.
More importantly, $\etat$ is positive, which is to be expected based on a
calculation for incompressible flows using SOCA, which is valid for $\Rm\ll1$.
To show this, one uses the fact that the Fourier transform of the velocity
correlation tensor is positive semidefinite.
We note in passing that this is not true for potential flows, for which
the turbulent diffusion tensor is negative semidefinite for $\Rm\ll1$;
see \cite{Rad11} for a recent demonstration using the TFM.

Returning now to the Roberts-IV flow, which is indeed incompressible, we
show in \Fig{petat} that $\etat$ changes sign from positive to negative
at $\Rm\approx4$.
This is clearly a result that cannot be recovered by SOCA.
A corresponding calculation for the turbulent passive scalar diffusivity,
$\kappat$ \citep[e.g.][]{BSV09}, shows that its value remains positive
and close to the SOCA value for small P\'eclet numbers, $\Pe=u_0/\kappa k_0$,
where $\kappa$ is the microphysical (molecular) passive scalar diffusivity.

Our TFM results show that, for $\Rm\approx8$, the total magnetic diffusivity,
$\eta+\etat(k,0)$ becomes negative, i.e., dynamo action by
the negative magnetic diffusivity effect is possible.
The critical value of $\Rm$ agrees with that found above through DNS.
The growth rate of the dynamo is given in implicit form
as a solution of the equation
\begin{equation}
\lambda(k)=-[\eta+\etat(k,\ii\lambda)]k^2
\label{lamimp}
\end{equation}
for $k=k_0$.
However, it is common to approximate $\etat(k_0,\omega)$ by $\etat(k_0,0)$,
and we refer to the corresponding solution as
\begin{equation}
\lambda(k)\approx\lambda_{\rm disp}(k)=-[\eta+\etat(k,0)]k^2,
\label{lamdisp}
\end{equation}
which is shown in the third panel of \Fig{petat} and compared with the
growth rate $\lambda_{\rm DNS}$ obtained by solving \Eq{dBdt} through DNS.
The agreement between $\lambda_{\rm disp}$ and $\lambda_{\rm DNS}$
is moderate and reminiscent of what has been found earlier \citep{HB09}.
By using test fields that grow exponentially at a rate that is equal
to the expected growth rate, $\lambda=\lambda_{\rm DNS}$, we find for
$\eta=0.020\,u_0/k_0$ (corresponding to $\Rm\approx50$), the value
$\etat(k_0,\ii\lambda_{\rm DNS})=-0.081\,u_0/k_0$ with
$\lambda_{\rm DNS}=0.061\,u_0k_0$, instead of the value
$\lambda=-0.070\,u_0k_0$ obtained with $\etat(k_0,0)=0$.
Thus, perfect agreement between DNS and TFM is obtained once the memory
effect included.

\begin{figure}\begin{center}
\includegraphics[width=\columnwidth]{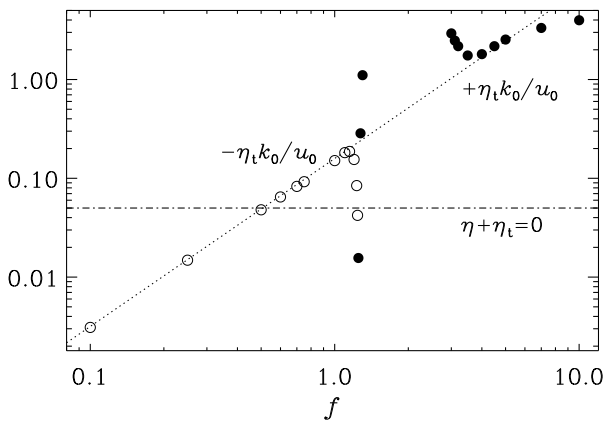}
\end{center}\caption[]{
Dependence of turbulent diffusivity on $f$ for $\Rm=20$ and $k=k_0$.
Negative (positive) values of $\etat$ are indicated with open (filled) symbols,
and the horizontal dash-dotted line indicates the region above which there
is dynamo action, because $\eta+\etat<0$.
The dotted line has a slope of 1.7 and is shown for orientation.
}\label{petat_fdep}\end{figure}

\subsection{Dependence on $f$}
Let us now discuss the dependence on the parameter $f$, which characterizes
the relative importance of vertical to horizontal motions.
We consider here the case of $\Rm=20$ and $k=k_0$.
The results are shown in \Fig{petat_fdep}.
The negative turbulent diffusivity dynamo is found to be
operating in the range $0.6\le f\le 1.23$, i.e., when the vertical
turbulent diffusivity is not much larger than the horizontal.

As indicated by the dotted line in \Fig{petat_fdep}, both for small
and for large values of $f$, there is an approximate power law dependence
with $|\etat|\sim f^{1.7}$.
However, in the range $1.3< f<3$ the TFM diverges and is unable to deliver
useful results.
Diverging results of the TFM are common and related to unstable eigenvalues
of the associated homogeneous system of equations solved in the TFM.
Usually, this problem can be avoided by restricting the analysis of
the test problems to limited time intervals
\citep{HDSKB09,RB10}, but in the present case the solutions were
diverging immediately.

\begin{figure}\begin{center}
\includegraphics[width=\columnwidth]{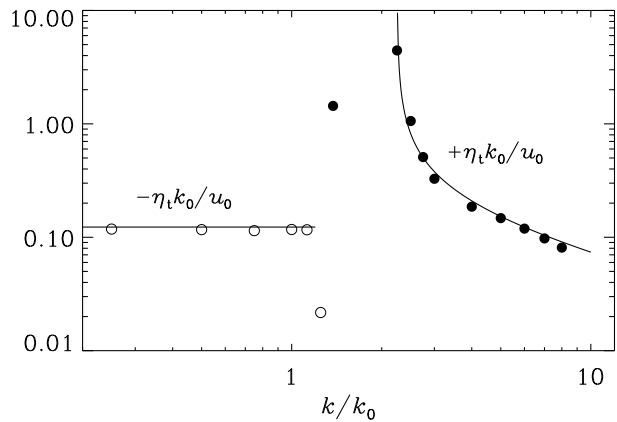}
\end{center}\caption[]{
Wavenumber dependence of turbulent diffusivity for $\Rm=20$ and $f=1$.
Note that dynamo action is only possible for $k/k_0\le1.25$,
i.e., when $\etat$ is negative.
}\label{petat_kdep}\end{figure}

\begin{figure}\begin{center}
\includegraphics[width=\columnwidth]{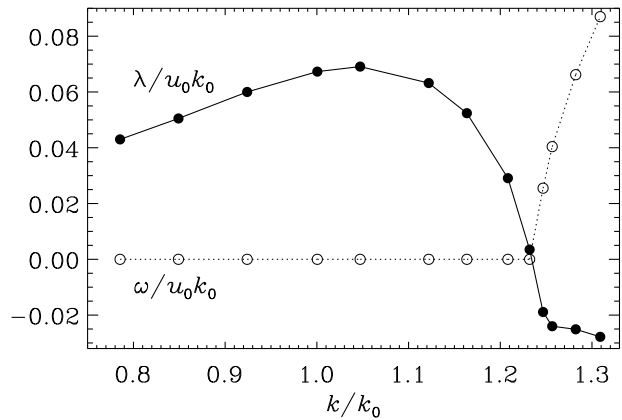}
\end{center}\caption[]{
Dependence of the growth rate $\lambda$ and the oscillation frequency $\omega$
on $k$ (using a correspondingly adjusted domain size $L_z=2\pi/k$)
for $\Rm=20$ and $f=1$.
In all cases with $\lambda>0$, we have $\omega=0$.
}\label{pDNS_Ldep}\end{figure}

\subsection{Scale dependence and memory effect}

Owing to the $k^2$ factor in \Eqs{lamimp}{lamdisp}, one might expect
dynamos driven by negative eddy diffusivity to grow
faster at larger values of $k$ (smaller scales), unless $\eta+\etat$
changes and becomes positive at larger $k$.
To study this, we now employ test fields with $k\neq k_0$.
In \Fig{petat_kdep} we show the $k$ dependence of $\etat(k,0)$.
It turns out that $\etat(k,0)$ is approximately constant for
$k\leq k_*\approx1.125\,k_0$, and positive with an approximate dependence
\begin{equation}
\etat(k,0)\approx{0.31\,u_0/k_0\over[(k-k_\infty)/k_0]^{0.7}}\quad
\mbox{for $k>k_\infty$},
\end{equation}
where $k_\infty\approx2.25\,k_0$.
In the range $k_*<k<k_\infty$ we have two data points in \Fig{petat_kdep}
that clearly deviate from the description above.
In addition, there are several other values of $k$ in this range where
the TFM again diverges and is unable to deliver useful results.

To illuminate the problem of intermediate $k$ values further,
we now use DNS to compute the growth rate as a function of the
domain size $L_z$, decreased according to $L_z=2\pi/k$.
The result is shown in \Fig{pDNS_Ldep}.
It turns out that $\lambda$ ($=\lambda_{\rm DNS}$)
has a maximum at $k/k_0\approx1.04$ (corresponding to $L_zk_0\approx6$).
Thus, large scale separation, as assumed in some analytic studies
\citep{LNVW99}, is neither needed nor necessarily helpful for the
operation of this negative eddy diffusivity dynamo.
Furthermore, for $k/k_0\approx1.23$ (corresponding to $L_zk_0\leq5.1$),
no dynamo is possible and the field decays in an oscillatory fashion.
The oscillation frequency $\omega$ grows sharply as $k$ increases
further; see the dotted line in \Fig{pDNS_Ldep}.
However, when the dynamo is excited, it is non-oscillatory.
While \cite{Rob72} also finds non-oscillatory behavior in the
dynamo cases, he finds non-oscillatory decaying solutions,
but at smaller $\Rm$.

At the level of a simplistic description of a mean-field
dynamo with negative eddy diffusivity, the occurrence of
oscillations in the subcritical case must be surprising.
However, this puzzle is easily resolved by reinstating the
$\omega$ dependence of $\etat(k,\omega)$ in \Eq{EMFkomega}.
This corresponds to the memory effect, i.e., the dependence of
$\meanEMF$ on the mean magnetic field at past times.

A simple prescription of the memory effect would be to assume
that $\etat$ is proportional to the analytic function
$1/(1-\ii\omega\tau)$, where
$\tau$ is a characteristic time scale of the flow.
Following \cite{HB09}, we replace $-\ii\omega$ by the Laplace
variable $s$ and assume that the growth rate is
equal to $\Rey\,s$ and the frequency is $\omega=-\mbox{Im}\,s$.
This leads to the dispersion relation
\begin{equation}
s\tau=-\tau\etat k^2/(1+s\tau)-\tau\eta k^2.
\end{equation}
Solving this quadratic equation for $s\tau$, we find
\begin{equation}
s_\pm\tau=-\half(1+n)\pm\half\sqrt{(1-n)^2-4n_{\rm t}},
\end{equation}
where $n=\tau\eta k^2$ and $n_{\rm t}=\tau\etat k^2$ have
been introduced.
Here, only the upper sign corresponds to physically realizable
solutions that can grow for negative eddy diffusivity, $\etat+\eta<0$.
In that case, $s\tau$ is real, but complex for positive
turbulent diffusivity, $n_{\rm t}>0$.
This explains qualitatively the occurrence of oscillatory decay,
except that this formula would also predict a narrow $n_{\rm t}$ interval
of non-oscillatory decay which is not seen in the data.

\begin{figure}\begin{center}
\includegraphics[width=\columnwidth]{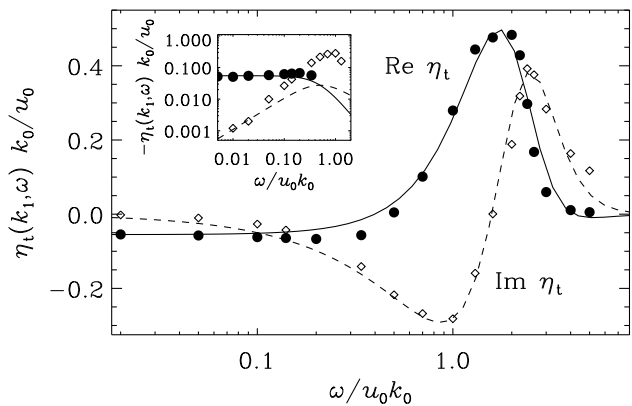}
\end{center}\caption[]{
Dependence of real and imaginary parts of $\tilde\etat$ on $\omega$
for the case $k/k_0\approx1.23$, $\Rm=20$ and $f=1$ compared with
the empirical fit given by \Eq{fit}.
Real (imaginary) parts are indicated by filled (open)
symbols and solid (dashed) lines.
The inset shows that for $\omega/u_0k_0<0.1$, the data are well
described by $\etat=-0.055\,u_0k_0/(1-\ii\omega\tau)$ with
$\tau\approx2/u_0k_0$.
}\label{po32}\end{figure}

The actual form of $\etat(k,\omega)$ near onset at
$k/k_0\approx1.23$ is of course more complicated.
The result, obtained using the method described by \cite{HB09},
is shown in \Fig{po32}.
For $\omega/u_0k_0>0.5$, a reasonable fit to the data is given by
\begin{equation}
\etat(k,\omega) \approx {u_0 k_0 \over [1+b(\omega/u_0k_0)^4]^2}
\sum_{n=0}^4 a_n\left(-{\ii\omega\over u_0k_0}\right)^n,
\label{fit}
\end{equation}
with empirical coefficients $a_0=-0.055$, $a_1=0.5$,
$a_2=-0.35$, $a_3=0.2$, $a_4=-0.02$, and $b=0.031$.
On the other hand, for small departures from the stationary state,
$\omega/u_0k_0\ll0.5$,
a good approximation is $\etat=-0.055\,u_0k_0/(1-\ii\omega\tau)$
with $\tau\approx2/u_0k_0$, confirming thus our initial ansatz.
For larger values of $k$, when $\Rey\,\etat$ becomes positive,
there is a rapid increase of $\tau$, which explains why the
aforementioned interval with non-oscillatory decay is absent.

\begin{figure}\begin{center}
\includegraphics[width=\columnwidth]{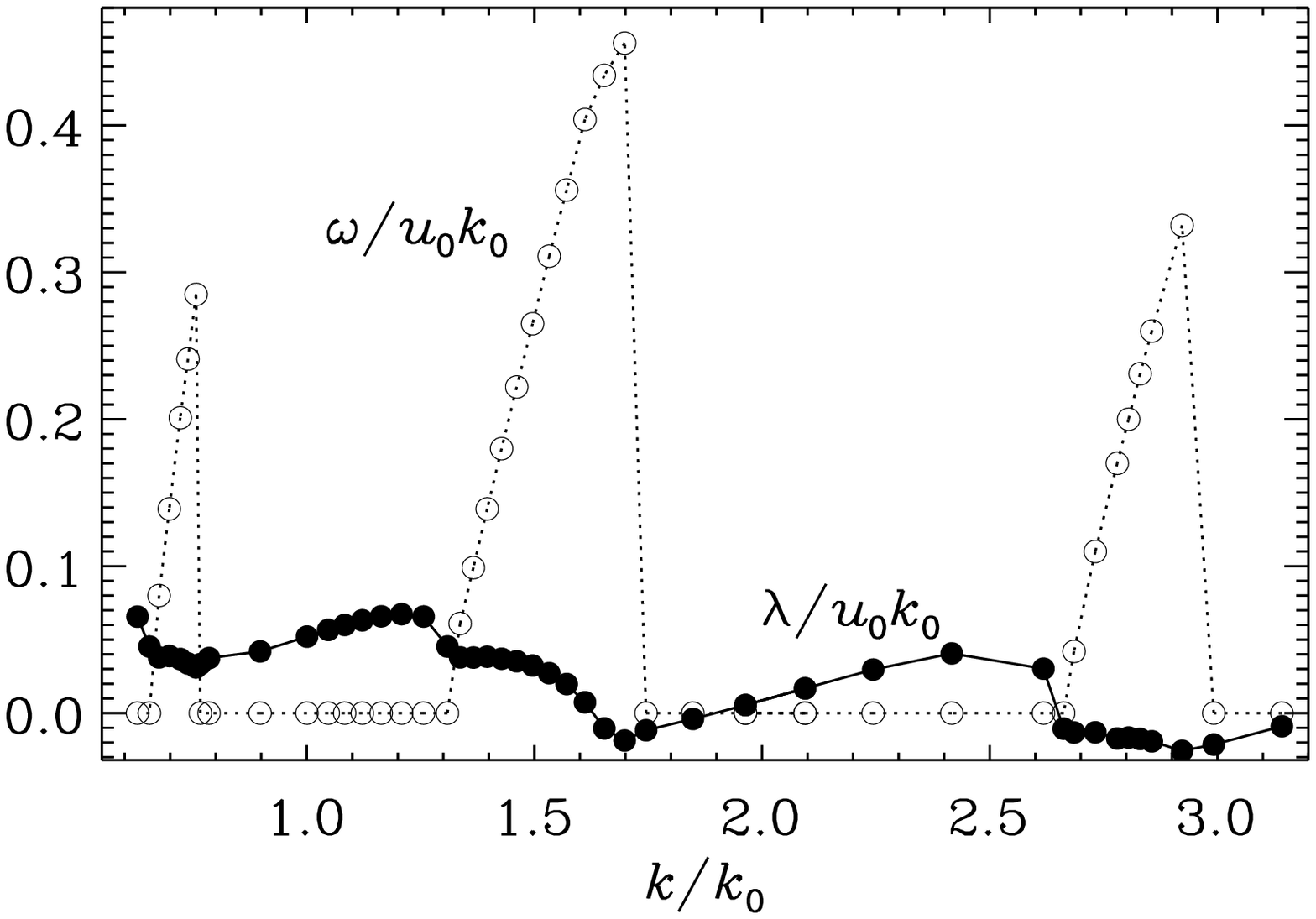}
\end{center}\caption[]{
Similar to \Fig{pDNS_Ldep}, but for $\Rm=100$.
Note the existence of growing oscillatory solutions in the ranges
$0.6\la k/k_0\la0.8$ and $1.3\la k/k_0\la1.7$.
}\label{pDNS_Ldep_Rm100}\end{figure}

In analogy with $\alpha$ effect mean-field dynamos, where the wavenumber
of the fastest growing mode increases with dynamo number, one may ask
whether this is also true of the negative magnetic diffusivity dynamo.
In \Fig{pDNS_Ldep_Rm100} we show the result for $\Rm=100$.
It turns out that the wavenumber of the fastest growing mode increases
slightly (from $k/k_0\approx1.04$ at $\Rm=20$ to $\approx1.21$ at $\Rm=100$).
For larger values of $k$, the solutions become again oscillatory, but,
in contrast to the case of smaller values of $\Rm$, the modes are now
not decaying.
Looking at \Fig{pDNS_Ldep_Rm100}, it becomes clear that the explanation
in terms of the simplest form of the memory effect no longer applies,
and that a more detailed dependence on $\omega$ would need to be considered.

\section{Modified Taylor--Green flows}

While the possibility of a dynamo driven by negative eddy diffusivity
has not been previously quantified for the Roberts-IV flow, it was discussed
in some detail by \cite{LNVW99} for the Taylor--Green (TG) flow,
\begin{equation}
\UU_{\rm TG}=u_0\pmatrix{
\sin k_0x\cos k_0y\cos k_0z\cr
-\cos k_0x\sin k_0y\cos k_0z\cr
0},
\end{equation}
and the modified TG flow, $\UU_{\rm TG}+A\UU_{\rm A}+B\UU_{\rm B}$,
where $A$ and $B$ denote the amplitudes of additional contributions
proportional to
$$
\UU_{\rm A}=u_0\pmatrix{
\sin2k_0x\cos2k_0z\cr
\sin2k_0y\cos2k_0z\cr
-(\cos2k_0x+\cos2k_0y)\sin2k_0z\cr
}
$$
and
$$
\UU_{\rm B}=u_0\!\pmatrix{
 (\sin k_0x\cos3k_0y+{5\over13}\sin3k_0x\cos k_0y)\cos k_0z\cr
-(\cos3k_0x\sin k_0y+{5\over13}\cos k_0x\sin3k_0y)\cos k_0z\cr
{2\over13}(\cos k_0x\cos3k_0y-\cos3k_0x\cos k_0y)\sin k_0z}\!,
$$
respectively.

\begin{figure}\begin{center}
\includegraphics[width=\columnwidth]{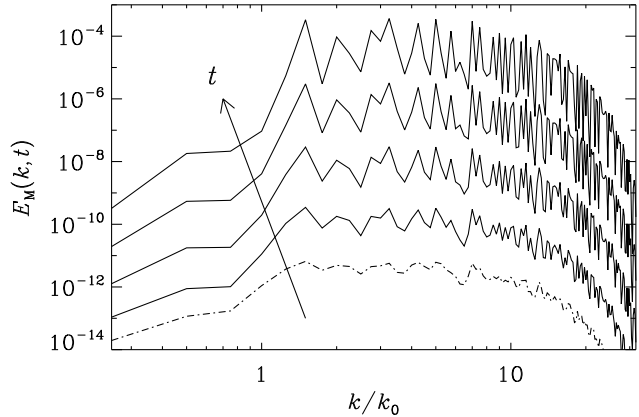}
\end{center}\caption[]{
Magnetic power spectra of DNS for the TG flow with $A=1$ and $B=0$
in a domain of size $L_x=L_y=L_z=8\pi/k_0$ using
$256^3$ mesh points and $\eta=0.02\,u_0/k_0$ at different times
during the exponential growth phase of the dynamo.
The earliest time is shown as a dash-dotted line.
}\label{pkt}\end{figure}

We have performed kinematic DNS with this flow and we indeed find dynamo action.
We carry out calculations for $\eta=0.02\,u_0/k_0$, which corresponds
to the case where the dynamo is mildly supercritical.
We consider the following three cases:
(a) A cube of size $L_x=L_y=L_z=2\pi/k_0$, using $128^3$ meshpoints;
(b) a cuboid with $L_z=4\,L_x$ and $L_y=L_x=2\pi/k_0$, using $128^2\times512$ meshpoints;
and (c) a cuboid with $L_x=L_y=4\,L_z$ and $L_z=2\pi/k_0$.
In all these cases the large-scale field obtained by averaging over the
$xy$ plane decays as a function of time, i.e., {\it no} large-scale dynamo
is obtained with this average.
However, there still remains, in principle, the possibility of a large-scale field developing that
is zero under $xy$ averaging but is non-zero under another averaging
procedure, e.g., Fourier filtering.
But even this possibility is ruled out because we observe
no growth of a large-scale field at $k\leq k_0$; see \Fig{pkt}.
This is found by calculating the spectrum of the magnetic field from simulations
with $256^3$ meshpoints and a domain size of $L_x=L_y=L_z=8\pi/k_0$.
In other words, a dynamo is observed but it is {\it not} a large-scale dynamo.
This is corroborated by the TFM which produces positive turbulent diffusivity
and vanishing $\alpha$ in all the aforementioned cases.
In case (a) $\etat=+0.135\,u_0/k_0$ for $A=1$ and $B=0$, and $\etat=+0.146\,u_0/k_0$ for
$A=1$ and $B=1$;
in case (b) $\etat=+0.158\,u_0/k_0$.
However, in case (c) the TFM becomes unstable.

In conclusion, the magnetic field structure of dynamos from the
modified Taylor--Green flows is quite different from that of dynamos
from the Roberts-IV flow.
In the latter, the mean fields contributed about 50\% to the total
field; see \Fig{ppxyaver}, while
for the former, most of the power occurred at small scales.
No significant mean magnetic field could thus be identified.

\section{Conclusions}

In the present work we have revisited the Roberts-IV flow using the TFM
to compute the full set of turbulent transport coefficients.
we confirm an earlier result of \cite{Til04} that a dynamo is possible
and that all components of the $\alpha$ tensor are vanishing.
In addition, we confirm the result of \cite{Rob72}
that there is a finite horizontally averaged mean
magnetic field, which should be explicable in terms of mean-field
dynamo theory.
The TFM reveals that the turbulent diffusivity tensor is isotropic in
the horizontal plane.
Moreover, in the regime where the dynamo is excited, the turbulent diffusivity
is sufficiently strongly negative such that the eddy (molecular plus turbulent)
diffusivity is negative.
This is an unusual situation in that the horizontal components of the mean
field are completely decoupled and grow independently with arbitrary
relative amplitudes and phase shifts, but the same growth rate.

Many laminar flows are only slow dynamos, i.e., the growth rate
goes to zero for large $\Rm$.
The Roberts-IV flow is no exception.
These dynamos are therefore not expected to be astrophysically relevant.
However, the method used to analyze such dynamos (TFM combined with DNS)
is now playing an important role in the study of astrophysical dynamos
for turbulent flows.
The present work highlights the accuracy of this method in that it enables
us to pinpoint the detailed nature of a dynamo exhibiting a finite averaged
magnetic field.

In the present case of laminar flow patterns, nonlocality is crucially
important.  In other words, turbulent transport is described by a
convolution of suitable integral kernels with
the mean fields in space and time rather than just
a multiplication.
The TFM is particularly well suited to deal with such cases.
For generic turbulent flows, as shown in earlier works by \cite{HB09} and \cite{RB12},
we expect these transport kernels to have a relatively simple form
and that complicated kernels, such as found here and in the earlier
work \citep{RB09} are atypical.
Note however, that even though most astrophysical flows are turbulent
and are expected to become statistically homogeneous and isotropic at
small scales; in practice large scale anisotropy and inhomogeneity
play an important role.
In many of those cases nonlocality cannot be neglected and
many Fourier modes need to be taken into account, as demonstrated
by \cite{Chatter} for flows driven by the magnetic buoyancy instability.

\section*{Acknowledgements}

We thank Alessandra Lanotte for inspiring us to look for
negative eddy diffusivity dynamos, and Matthias Rheinhardt,
Karl-Heinz R\"adler, and the referee for detailed
comments regarding our paper.
Financial support from the Scientific \& Technological Research Council
of Turkey (T\"UB\.ITAK) and the European Research Council under the AstroDyn
Research Project 227952 are gratefully acknowledged.
The computations have been carried out at the National Supercomputer
Centre in Ume{\aa} and at the Center for Parallel Computers at the
Royal Institute of Technology in Sweden.


\vfill\bigskip\noindent\tiny\begin{verbatim}
$Header: /var/cvs/brandenb/tex/ebru/tilgner/paper.tex,v 1.63 2013-04-04 15:49:39 ebru Exp $
\end{verbatim}

\end{document}